# Materials considerations for forming the topological insulator phase in InAs/GaSb heterostructures

B. Shojaei[1,2], A. P. McFadden[3], M. Pendharkar[3], J. S. Lee[2], M.E. Flatté[4], and C. J. Palmstrøm[1, 2, 3, *]

1. Materials Department, University of California, Santa Barbara, CA 93106, USA
2. California NanoSystems Institute, University of California, Santa Barbara, CA 93106, USA
3. Department of Electrical and Computer Engineering, University of California, Santa Barbara, CA 93106, USA
4. Optical Science and Technology Center and Department of Physics and Astronomy, University of Iowa, Iowa City, Iowa 52242, USA

*Abstract*

In an ideal InAs/GaSb bilayer of appropriate dimension in-plane electron and hole bands overlap and hybridize, and a topologically non-trivial, or quantum spin Hall (QSH) insulator, phase is predicted to exist[1]. The in-plane dispersion's potential landscape, however, is subject to microscopic perturbations originating from material imperfections. In this work, the effect of disorder on the electronic structure of InAs/GaSb bilayers was studied by the temperature and magnetic field dependence of the resistance of a dual-gated heterostructures gate-tuned through the inverted to normal gap regimes. Conduction in the inverted (predicted topological) regime was qualitatively similar to behavior in a disordered two-dimensional system. The impact of charged impurities and interface roughness on the formation of topologically protected edge states and an insulating bulk was estimated. The experimental evidence and estimates of disorder in the potential landscape indicated the potential fluctuations in state-of-the-art films are sufficiently strong such that conduction in the predicted topological insulator (TI) regime was dominated by a symplectic metal phase rather than a TI phase. The implications are that future efforts must address disorder in this system and focus must be placed on the reduction of defects and disorder in these heterostructures if a TI regime is to be achieved.

___________________________

*Electronic mail: cpalmstrom@ece.ucsb.edu

*Manuscript*

The prediction of the existence of a topologically non-trivial, or quantum spin Hall (QSH) insulator, phase in InAs/GaSb heterostructures[1] has significant implications for fault-tolerant quantum information processing[2]. The prospect of implementing quantum computing architectures on technologically mature III-V requires topologically protected helical edge channels and an insulating bulk in InAs/GaSb heterostructures.

Evidence supporting the existence of helical edge states has been found in observations of conductance quantized near predicted values for single mode conduction and measured in both local and non-local configurations[3–6]. In some of these experiments silicon or beryllium doping was required to suppress bulk conductivity[5,6]. These experiments confirm conduction close to quantized values over micron-sized device geometries but do not incontrovertibly confirm helicity. Furthermore, they report suppressed bulk conduction, but do not confirm if the bilayer structure was tuned to the inverted gap regime. Additional evidence supporting the existence of helical edge states is lacking. In particular, reports on the temperature dependence of the conductivity when the Fermi level was interpreted to be in the hybridization gap showed it to be stagnant with changes in temperature[4–6] or to increase with increasing temperature[7], and are in disagreement with theoretical expectations for conduction through helical edge states[8].

This work reports on the temperature dependence of the resistance of a dual-gated 11 nm InAs / 8 nm GaSb heterostructure gate-tuned through the inverted to normal gap regimes. The inverted and normal regimes were identified by measurements of the evolution of resistance as a function of the two gates and under in-plane and out-of-plane magnetic fields. The resistance maxima monotonically decreased with increasing temperature over the entire phase diagram. A large positive magnetoresistance was observed when the structure was gate-tuned to the inverted regime. Conduction in the inverted (predicted TI) regime was qualitatively similar to behavior in a disordered two-dimensional system. The impact of charged impurities and interface roughness on the formation of topologically protected edge states and an insulating bulk were considered. Potential fluctuations in the electronic band structure for realistic levels of charged disorder were calculated using a gated heterostructure model. **k·p** calculations were used to estimate the effect of variations in film thickness over space on the electronic structure. The experimental evidence and estimates of disorder in the potential landscape indicate the potential fluctuations in state-of-the-art films are sufficiently strong such that conduction in the predicted topological insulator (TI) regime will be dominated by a symplectic metal phase rather than a TI phase.

The 11 nm InAs / 8 nm GaSb bilayer heterostructure was grown by molecular beam epitaxy on a GaSb:Te (001) substrate. A schematic of the heterostructure, with integrated 50 nm $Al_2O_3$ gate dielectric and Ti/Au gate metallization is shown in Fig. 1(a). The bilayer heterostructure is surrounded by a 50 nm AlSb top barrier and a 30 nm AlSb bottom barrier, a 10 period 2.5 nm AlSb / 2.5 nm GaSb superlattice and a 100 nm $AlAs_{0.1}Sb_{0.9}$ electrically insulating buffer. This dual-gated device structure has been previously shown to linearly vary the electron density in single InAs quantum wells with both the Ti/Au front gate, $V_f$, and the GaSb:Te back gate, $V_b$ over a wide gate voltage range[9]. A **k·p** calculation of the in-plane dispersion of a coupled electron-hole quantum well representing the heterostructure under study is shown in Fig. 1(b). Details of the computational approach are available in Refs 10 and 11[10,11]. The hybridization gap is calculated to be 0.8 meV. A dual gated Hall bar device fabricated over the heterostructure is shown in Fig. 1(c). The dual gated device allows for control of the subband levels, enabling the electronic structure to be tuned from a normal gap dispersion where the minimum of $E_1$ is higher in energy than the maximum of $H_1$ to an inverted gap dispersion where the minimum of $E_1$ is lower in energy than the maximum of $H_1$, while independently controlling the Fermi level[12,13].

Measurements were performed using standard lock-in techniques under an excitation current of 1 nA and 10 nA and at cryogenic temperatures in an adiabatic demagnetization refrigerator with a base temperature of 50 mK, a He-4 cryostat with a base temperature of 1.8 K, and a dilution refrigerator equipped with a sample tilter with a base temperature of 20 mK. In all systems, the device was cooled under zero gate bias.

The dependence of the longitudinal resistance on the front and back gate voltages at a sample temperature of 165 mK is shown in Fig. 2(a). The resistance map is divided into several regions surrounding and including a line of maximum resistance which itself observes a minimum value of approximately 7 k$\Omega$ at $V_b$ ~ 0.35 V, $V_f$ ~ -0.55 V. The (+) near the center of the map indicates the origin in gate space. Longitudinal and Hall magnetoresistance measurements, as shown for zero gate bias in Figs. 2(b) and 2(c), were used to determine the evolution of carrier density as a function of $V_f$ and $V_b$. The channel resistance was measured by modulating the front gate at fixed back gate. The hysteresis in modulating the front gate is shown in Fig. 2(d). The current dependence of the channel resistance maximum over a front gate sweep at a fixed back gate of -0.5 V is shown in Fig. 2(e).

Longitudinal and Hall magnetoresistance measurements identified transport in region I in Fig. 2(a) as being electron dominated with electron mobility exceeding 200,000 cm$^2$/Vs. Region II showed mixed conduction with majority n-type carriers. Magnetoresistance was measured at twenty-one points along line A, approximating a line of constant electric field. A subset of the measurements is shown in Figs. 3(a) and 3(b), and the longitudinal resistance, extracted carrier density and carrier type are shown in Fig. 3(c). Conduction in region V was of mixed type but through majority p-type carriers. Conduction in region VI was dominated by holes with mobility exceeding 20,000 cm$^2$/Vs.

Measurements were performed under in-plane magnetic field perpendicular to the current path of the Hall bar at a sample temperature of 165 mK. The dependence of the resistance along line A at 0 T and 1 T in-plane magnetic field is shown in Fig. 3(d). The resistance was found to decrease as a function of magnetic field along line A, and identifies region III as having an inverted subband structure[14,15]. The same measurement was performed across region IV with gate voltage modulated using a similar slope as that of line A, and is shown in Fig. 3(e). The relative change in the resistance as a function of in-plane magnetic field was significantly less over region IV and suggests the presence of a normal gap subband structure over region IV. The Hall voltage as a function of in-plane magnetic field was used to determine a misalignment of the in-plane magnetic field of approximately 2 degrees. Similar in-plane magnetic field measurements were performed in a dilution refrigerator equipped with a sample tilter where near perfect in-plane field alignment could be achieved. In this configuration the resistance maximum measured over one line across region III was found to monotonically decrease from 0 T to 4 T in 0.25 T increments. The resistance maximum measured over one line across region IV slightly increased over the same magnetic field range. The resistance minimum along the line of maximum resistance at $V_b \sim 0.35$ V, $V_f \sim -0.55$ V is interpreted to be due to a gap closing in the subband structure.

An approximation of the size of the hybridization gap was obtained by the ratio of the change in density across the gap over the density of states, and a range of $E_g = \frac{\pi \hbar^2 \cdot \Delta n}{m^*} \approx$ 3 meV $-$ 9 meV was determined, where $\Delta n$ is estimated to be $5 \times 10^{10}$ cm$^{-2}$ to $1.6 \times 10^{11}$ cm$^{-2}$ by extrapolating the carrier density and the associated error in its measurement far from the gap to points 11 and 15 along line A and straddling the gap. An effective mass is estimated as $m^* = 0.04 m_e$ from the temperature dependence of Shubnikov de Haas oscillations measured in region II up to a sample temperature of 20 K.

The resistance maxima in the space of gate voltages yielding a hybridized band structure and with the Fermi level tuned to the hybridized gap are of the order 8 kΩ. This value is far below the expected value of several multiples of $\frac{h}{e^2}$ for conduction in long helical edges with a phase coherence length smaller than the physical edge length. The temperature dependence of the resistance maxima of the longitudinal resistance measured by modulation of $V_f$ at several fixed $V_b$ is shown in Fig. 4(a). For Vb = -0.7 V and $V_b$ = 0.2 V, the dispersion is hybridized. For $V_b$ equal to and greater than 0.4 V, the conduction is measured through a normal gap. Figure 4(b) shows the temperature dependence at several additional back gate voltages extending further into the normal gap.

The dependence of resistance on temperature for conduction measured in the hybridization gap observes a saturating character at low temperature. This behavior and the large positive magnetoresistance for conduction through the hybridization gap is in qualitative agreement with conduction in a disordered two-fluid (electron-hole) system[16]. The temperature dependence is not in agreement with the predictions that consider helical edge channels coupled to charge puddles modeled as quantum dots[8]. The temperature dependence of the resistance for conduction measured in the normal gap observes an activated behavior but also appears to saturate at low temperatures. The saturation in the normal gap regime may be due to anomalous edge channels.[17]

The origin of electron and hole charge fluctuations in an InAs/GaSb heterostructure would result from disorder of sufficient strength to create potential fluctuations that force the conduction band below and the valence band above the Fermi level randomly over space. A simple model was used to estimate the strength of the potential fluctuations in the heterostructure. In this model, depicted in the inset of Fig. 5, a layer of randomly distributed donor impurities are placed at the III-V/gate dielectric interface for the heterostructure, where $l_g$ is the distance between the gate and bilayer, and $l_d$ is the distance between the donor layer and the bilayer. For a donor layer of sheet charge density $n_d$, Tripathi and Kennett developed an analytical solution for the resulting potential fluctuations given the above model and accounting for screening by carriers in the two-dimensional system:[18]

$$e\sqrt{\langle\delta\phi^2\rangle} = \frac{\sqrt{n_d}e^2}{4\sqrt{\pi}\epsilon} ln\left(1 + \left(\frac{R_c}{l_d}\right)^2\right)^{1/2},$$

where $R_c = \sqrt{n_d}/\sqrt{\pi}n_e$. The carrier density in the heterostructure was ~4 ($10^{11}$ cm$^{-2}$) at zero gate bias; therefore, $n_d$ can be assumed at least this large, and studies of the well carrier density as a function of the top barrier thickness[19] and the gating efficiency in similar devices[20] suggest it is in the range of $10^{12} - 10^{13}$ cm$^{-2}$. The calculated mean potential fluctuation as a function of $n_d$ for two different carrier densities, $n_e$, is depicted in Fig. 5. The dielectric constant, $\epsilon$, was assumed to be $10\epsilon_o$. The mobility of electrons confined to InAs for the same model calculated following the treatment of Stern and Howard[21], assuming a square well wavefunction[22] and a two-dimensional carrier density $n_e = 1\times10^{11}$ cm$^{-2}$ is also depicted in Fig. 5. The calculated mean potential fluctuation is substantially larger than the size of the hybridization gap measured in the heterostructure over the range of $n_d$.

An estimate of the mobility for the above model would be of the order $10^6$ cm$^2$/Vs. The measured mobility suggests the estimated potential fluctuations are a conservative estimate, and that other factors such as charge impurities in closer proximity to the bilayer and short range disorder such as interface roughness and alloy disorder should be considered.

Potential fluctuations originating from variations in film thicknesses over space, or interface roughness, were estimated by **k·p** calculations. The in-plane [100] dispersion of the electronic states forming the hybridization gap of three InAs / GaSb bilayers are shown in Figs. 6 (a)-(c). The variations in film thickness result in a shift of the electronic states relative to the vacuum level. When bilayer dimensions are varied +/- 1 Å in each component, as shown in Figs. 6(a) and 6(c), the hybridization gap no longer overlaps with the gap of an unmodified bilayer with dimensions 11.0 nm InAs and 8.0 nm GaSb , as shown in Fig. 6(b). The effect is further illustrated in Fig. 6(d), which depicts a calculation of the band edges of the hybridization gap relative to the vacuum level for changes in film thickness. The level of interface roughness that causes gap misalignments over space is well within the limits of that found in InAs/GaSb/AlSb heterostructures[23,24].

The implication of this work is that even in a structure with mobility substantially higher than in silicon doped heterostructures wherein quantized conductance has been reported[5], the level of disorder is sufficiently high to suppress the topological phase[25] and lead to conduction through the bulk. Improvements in material quality are one path toward reducing potential fluctuations due to disorder. It is noted that in InAs quantum wells negative persistence photoconductivity and an associated enhancement in the quantum lifetime under illumination was observed by Lo and coworkers[26], and may be a technique to reduce charge disorder in InAs/GaSb heterostructures.

This work reported on the temperature and magnetic field dependence of the resistance of a dual-gated 11 nm InAs / 8 nm GaSb heterostructure gate-tuned through the inverted to normal gap regimes. The resistance maxima monotonically decreased with increasing temperature over the entire phase diagram. A large positive magnetoresistance was observed when the structure was gate-tuned to the inverted regime. Conduction in the predicted TI regime was qualitatively similar to behavior in a disordered two-dimensional system. Potential fluctuations were estimated to be sufficiently strong such that conduction in the predicted TI regime was likely dominated by a symplectic metal phase rather than a TI phase. The implications are that future efforts must address disorder in this system and focus must be placed on the reduction of defects and disorder in these heterostructures if a TI regime is to be achieved.

*Acknowledgements*

This work was supported by Microsoft Research Station Q. The authors thank C. Nayak, R. M. Lutchyn, L. I. Glazman and M. H. Freedman for useful discussions. This work made use of the central facilities of the UCSB MRL, which is supported by the MRSEC program of the National Science Foundation under Award No. DMR-1121053. This work also made use of the UCSB Nanofabrication Facility, a part of the NSF funded NNIN network, and of the California NanoSystems Institute. A part of the measurements were performed at the National High Magnetic Field Laboratory in Tallahassee, Florida.

*Author contributions*

BS and CJP designed the experiment. BS designed the heterostructure. BS and MP grew the heterostructure by molecular beam epitaxy. BS fabricated the devices. BS, APM, MP and JSL performed the low temperature measurements. BS analyzed the data. BS calculated the estimates of potential fluctuations due to ionized impurities. MEF performed and analyzed the **k·p** calculations. CJP supervised the project. All authors contributed to writing the manuscript.

*Figures*

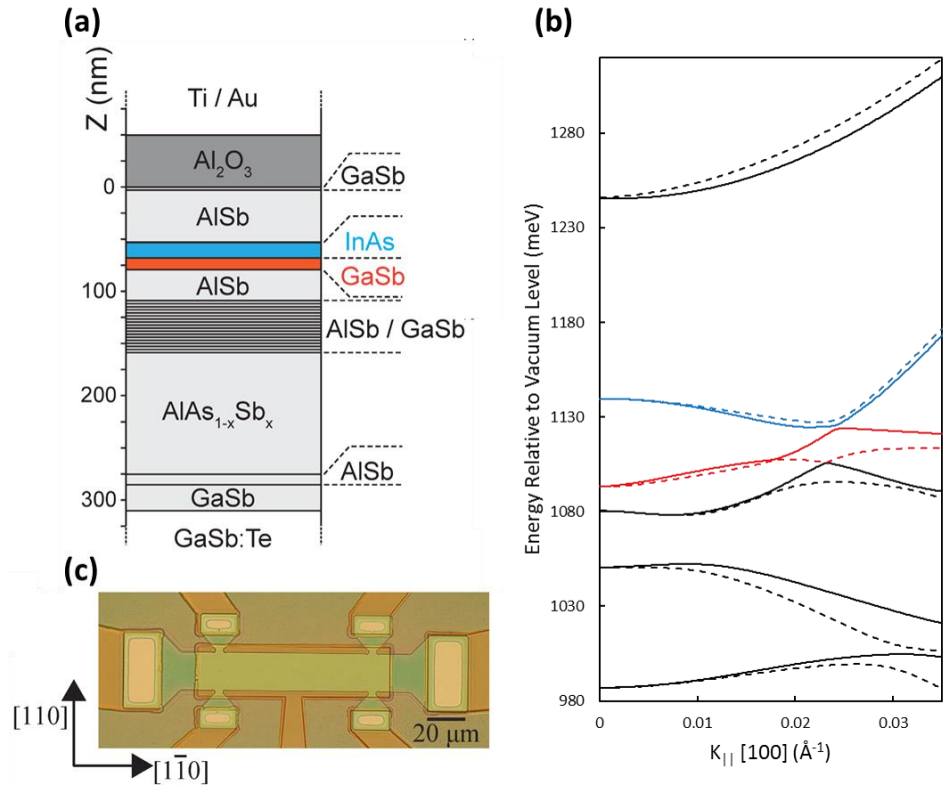

**Figure 1.** (a) A schematic of the 11 nm InAs / 8 nm GaSb bilayer heterostructure with $Al_2O_3$ gate dielectric and Ti/Au front gate. (b) The calculated in-plane [100] dispersion calculated for the 11 nm InAs / 8 nm GaSb bilayer. (c) An optical micrograph of the 80 x 20 μm² Hall bar device used in this study prior to gate and ohmic metallization. The long axis of the hall bar is oriented along the [1-10] crystallographic direction.

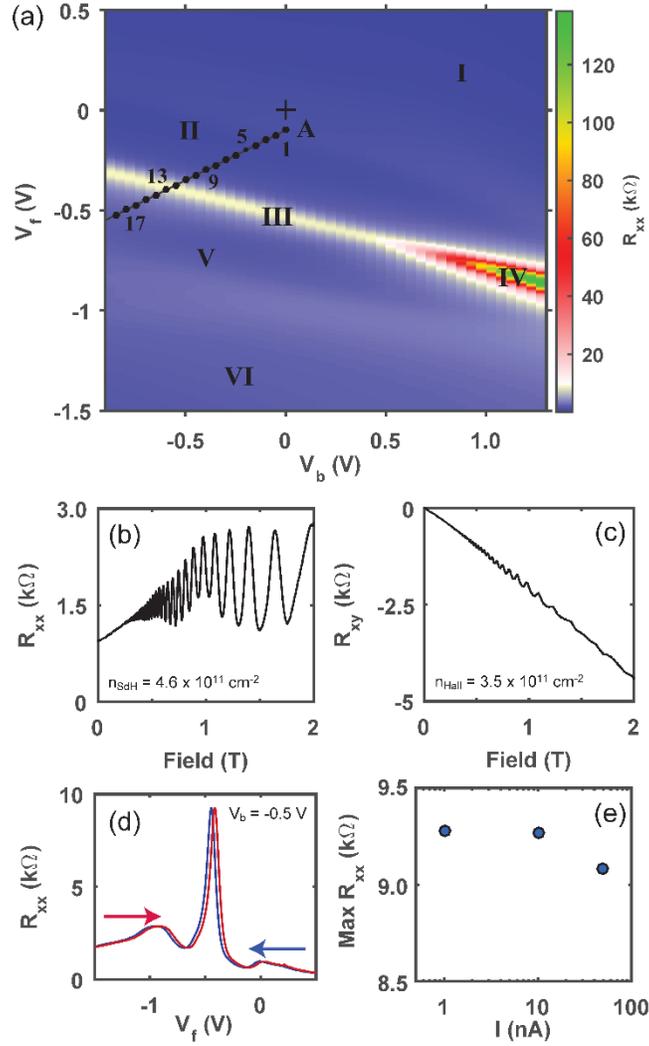

**Figure 2.** (a) The dependence of longitudinal resistance on front and back gate voltages at a sample temperature of 165 mK. Line A, spanning regions II and V, follows an estimate of constant electric field over the InAs/GaSb heterostructure across the inverted gap. (b) Magnetoresistance measured at zero gate bias. (c) Hall magnetoresistance measured at zero gate bias. (d) Longitudinal resistance as a function for front gate at a fixed back gate of -0.5 V for decreasing and increasing front gate voltage sweeps. (e) The maximum value of the longitudinal resistance as a function of excitation current measured as the front gate is modulated as in (d) at a fixed back gate -0.5 V.

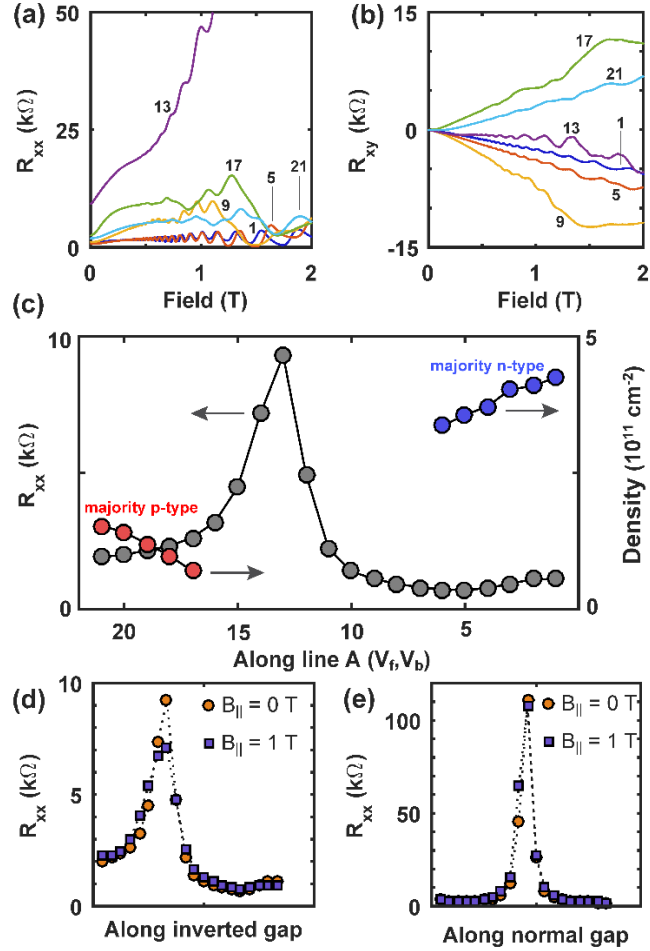

**Figure 3.** (a) The longitudinal magnetoresistance measured along several points along the line A. (b) The Hall magnetoresistance measured along several points along line A. (c) Longitudinal resistance and carrier density measured along line A. (d) Longitudinal resistance along line A at 0 T and 1 T in-plane magnetic fields. (e) Longitudinal resistance along a line approximately parallel to the line A but crossing the normal gap regime (region IV) and at 0 T and 1 T in-plane magnetic fields.

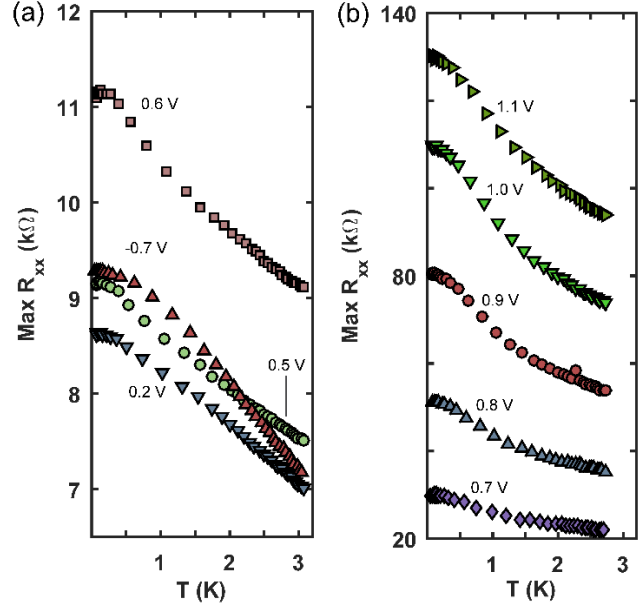

**Figure 4.** (a) Temperature dependence of the resistance maximum of the longitudinal resistance over front gate modulation at several fixed back gate voltages. (b) Temperature dependence of the resistance maximum of the longitudinal resistance over front gate modulation at several additional fixed back gate voltages.

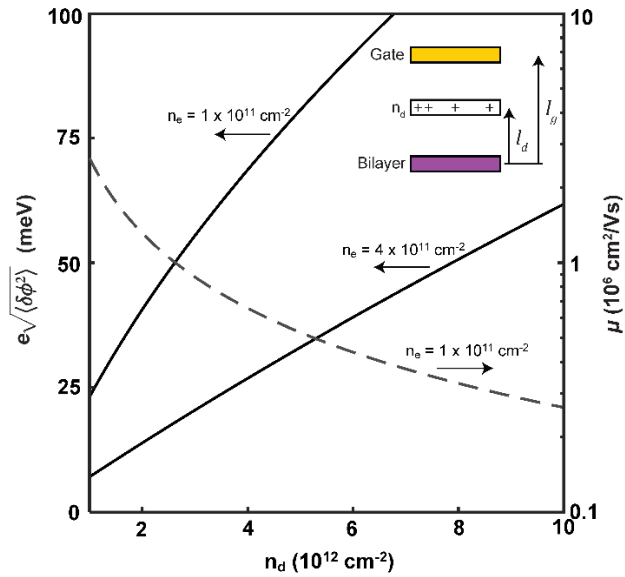

**Figure 5.** The calculated mean potential fluctuation for bilayer densities, $n_e = 1\times10^{11}$ cm$^{-2}$ and $n_e = 4\times10^{11}$ cm$^{-2}$, and the calculated mobility for a bilayer density, $n_e = 1\times10^{11}$ cm$^{-2}$, as a function of a two-dimensional ionized donor layer for the gated heterostructure model shown in the inset.

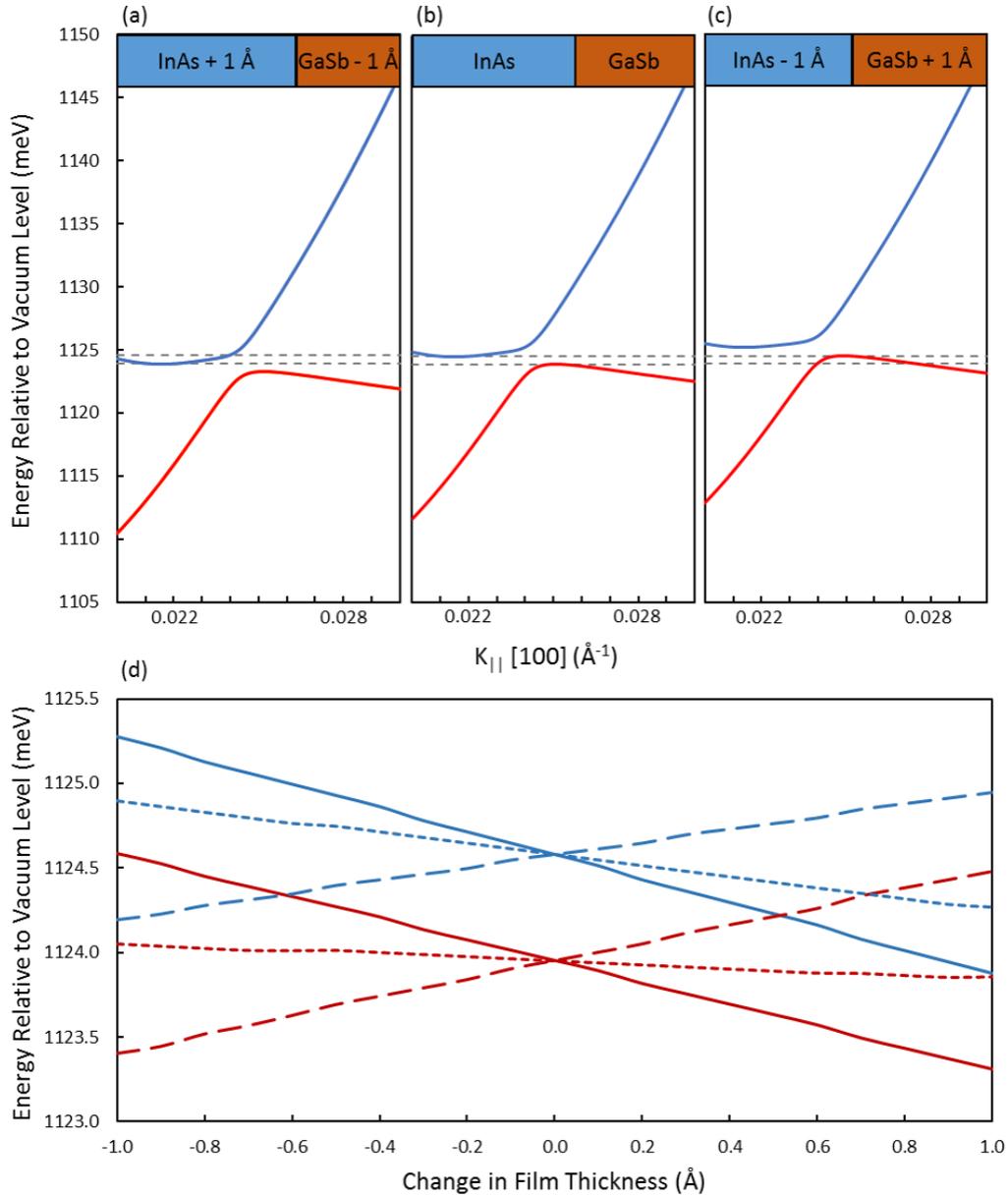

**Figure 6.** (a) The in-plane [100] dispersion of an 11.1 nm InAs / 7.9 nm GaSb bilayer near the hybridization gap. (b) The in-plane [100] dispersion of an 11.0 nm InAs / 8.0 nm GaSb bilayer near the hybridization gap. (c) The in-plane [100] dispersion of an 10.9 nm InAs / 8.1 nm GaSb bilayer near the hybridization gap. The grey dashed lines indicate the band edges of an 11.0 nm InAs / 8.0 nm GaSb bilayer relative to the vacuum level. (d) Band edges of the hybridization gap relative to the vacuum level for a change in the InAs film thickness with corresponding change in GaSb film thickness of the same magnitude with opposite sign (solid lines), for a change in the GaSb film thickness (long dashes), and for a change in the InAs film thickness (short dashes). The origin in film thickness corresponds to an 11 nm InA / 8 nm GaSb bilayer.